\documentclass{ws-procs9x6}

\begin{document}

\title{The static Penta-quark Potential \\ in Lattice QCD\footnote{
\uppercase{T}he lattice \uppercase{QCD} 
simulation has been done on \uppercase{NEC-SX5} at \uppercase{O}saka \uppercase{U}niversity.}
\vspace{-0.4cm}
}

\author{Fumiko Okiharu}

\address{Nihon University, 
1-8-14 Kanda Surugadai, Chiyoda, Tokyo 101-8308, Japan}

\author{
\vspace{-0.2cm}
Hideo Suganuma}

\address{Tokyo Institute of Technology, 
Ohokayama, Meguro, Tokyo 152-8551, Japan}

\author{
\vspace{-0.2cm}
Toru T. Takahashi}

\address{YITP, Kyoto University, 
Kitashirakawa, Sakyo, Kyoto 606-8502, Japan}

\maketitle

\abstracts{
\vspace{-0.3cm}
We perform the first study 
for the static penta-quark (5Q) potential 
in lattice QCD with $\beta$=6.0 and $16^3 \times 32$ at the quenched level. 
Accurate results of the 5Q potential are extracted from the 5Q Wilson loop  
using the smearing method, which enhances the ground-state component. 
The tetra-quark potential for the $\rm QQ$-$\rm {\bar Q}{\bar Q}$ system is also studied in lattice QCD.
The multi-quark potentials are found to be well described as a sum of the one-gluon-exchange Coulomb term and 
the multi-Y linear confinement term based on the flux-tube picture. 
\vspace{-0.3cm}
}

\section{Introduction}

The recent experimental discoveries of $\Theta^+(1540)$,\cite{Theta1540,Z04} 
$\Xi^{--}$(1862)\cite{NA49} and $\Theta_c$(3099)\cite{H1}
as the candidates of penta-quark (5Q) baryons 
are expected to reveal new aspects of QCD and hadrons. 
Such experiments were motivated by the theoretical prediction by Diakonov {\it et al.}\cite{DPP97}, and 
many theoretical studies have been done 
to clarify the 5Q baryons,$^{2,6-17}$
However, there are so many open problems for 
several remarkable features of the 5Q baryons. 
Experimental data indicate extremely narrow decay widths
and small masses of the 5Q baryons, and the parity determination is also 
an open problem. 
For the physical understanding of these features, 
theoretical analyses are necessary as well as the experimental studies. 
In particular, to clarify the inter-quark force in the multi-quark 
system based on QCD is required for the realistic modeling of the multi-quark system. 

For this purpose, we perform the static penta-quark potential 
$V_{\rm 5Q}$ in lattice QCD.
We here investigate QQ-$\rm \bar{Q}$-QQ type configurations 
for the 5Q system\cite{OST04,STOI04} where 
the two QQ clusters belong to $\bf 3^*$ representation in $\rm SU(3)_c$, 
since the $\bf 3^*$ diquark has a smaller energy than the $\bf 6$ diquark.\cite{JW03}
Similarly, we investigate QQ-$\rm \bar{Q}\bar{Q}$ type configurations 
for the 4Q system where the QQ and the $\rm \bar{Q}\bar{Q}$ 
clusters belong to $\bf 3^*$ and $\bf 3$ representation, respectively.

\section{Theoretical Ansatz: OGE Coulomb plus multi-Y Ansatz}

To begin with, we give a theoretical consideration for the multi-quark potential. 
From recent lattice QCD studies, 
the static three-quark (3Q) potential for baryons is found to obey 
the Coulomb plus Y-type linear potential, {\it i.e.}, the Y-Ansatz.\cite{TS} 
In fact, the lattice data for the $\rm Q\bar{Q}$ and the 3Q potentials 
can be well described as    
\begin{eqnarray}
V_{\rm Q \bar{Q}}&=&-\frac{A_{\rm Q \bar{Q}}}{r}
+\sigma_{\rm Q \bar{Q}} r+C_{\rm Q \bar{Q}}, \\ 
V_{\rm 3Q}&=&-A_{\rm 3Q}\sum_{i<j}\frac1{|{\bf r}_i-{\bf r}_j|}
+\sigma_{\rm 3Q} L_{\rm min}+C_{\rm 3Q}, 
\end{eqnarray}
where 
$A$ denotes the Coulomb coefficient, 
$\sigma$ the string tension, $C$ a constant and 
$L_{\rm min}$ the minimal length of the flux tube 
linking the valence quarks. 
These lattice QCD results indicate the flux-tube picture for hadrons. 

We generalize the Y-Ansatz  to the static 5Q potential $V_{\rm 5Q}$, and conjecture 
the one-gluon-exchange (OGE) Coulomb plus multi-Y linear potential\cite{OST04,STOI04} 
for $V_{\rm 5Q}$, {\it i.e.}, OGE Coulomb plus multi-Y Ansatz, 
\begin{eqnarray}
V_{\rm 5Q}
&=&\frac{g^2}{4\pi} \sum_{i<j} \frac{T^a_i T^a_j}{|{\bf r}_i-{\bf r}_j|}
+\sigma_{\rm 5Q} L_{\rm min}+C_{\rm 5Q} \nonumber \\
&=&-A_{\rm 5Q}\{ ( \frac1{r_{12}}  + \frac1{r_{34}}) 
+\frac12(\frac1{r_{15}} +\frac1{r_{25}} +\frac1{r_{35}} +\frac1{r_{45}}) 
\nonumber \\
& &+\frac14(\frac1{r_{13}} +\frac1{r_{14}} +\frac1{r_{23}} +\frac1{r_{24}}) \}
+\sigma_{\rm 5Q} L_{\rm min}+C_{\rm 5Q},
\label{theorform}
\end{eqnarray}
where the two quarks at (${\bf r}_1$, ${\bf r}_2$) and those at (${\bf r}_3$, ${\bf r}_4$) belong to $\bf 3^*$ representation, respectively, 
and the antiquark locates at ${\bf r}_5$. 
Here, $L_{\rm min}$ is defined as the minimal length of the flux tube 
linking all the valence (anti-)quarks. 

For the 4Q system, OGE Coulomb plus multi-Y Ansatz is expressed as  
\begin{equation}
V_{\rm 4Q}=-A_{\rm 4Q}\{(\frac1{r_{12}}+\frac1{r_{34}})
+\frac1{2}(\frac1{r_{13}}+\frac1{r_{14}}+\frac1{r_{23}}+\frac1{r_{24}})\}
+\sigma_{\rm 4Q}L_{\rm min}+C_{\rm 4Q}, 
\label{theorform4Q}
\end{equation}
where two quarks locate at (${\bf r}_1$, ${\bf r}_2$) and two antiquarks at (${\bf r}_3$, ${\bf r}_4$). 

We theoretically expect $\frac12 A_{\rm Q \bar Q}=A_{\rm 3Q}=A_{\rm 4Q}=A_{\rm 5Q}$ as the OGE result, 
and $\sigma_{\rm Q \bar Q}=\sigma_{\rm 3Q}=\sigma_{\rm 4Q}=\sigma_{\rm 5Q}$ as the universality of the string tension.

\section{Multi-quark Wilson loops and multi-quark potentials}

In QCD, the static potential $V$ is derived from the Wilson loop $W$ 
as 
\begin{eqnarray}
V=-\lim_{T\rightarrow \infty}\frac1{T}{{\rm ln} \langle W \rangle}.
\end{eqnarray}
The static multi-quark potentials can be also obtained from the corresponding multi-quark Wilson loops.  
As shown in Figs.1 and 2, we define the 5Q Wilson loop $W_{\rm 5Q}$\cite{OST04} 
and the 4Q Wilson loop $W_{\rm 4Q}$ as  
\begin{eqnarray}
W_{\rm 5Q}&\equiv& \frac1{3!} \epsilon^{abc} \epsilon^{a'b'c'}
\tilde M^{aa'}
(\tilde L_3\tilde L_{12}\tilde L_4)^{bb'}
(\tilde R_3\tilde R_{12}\tilde R_4)^{cc'}, \\
W_{\rm 4Q}&\equiv& \frac1{3} {\rm tr}
(\tilde M_1\tilde L_{12}\tilde M_2\tilde R_{12}),
\end{eqnarray}
where $\tilde L_i, \tilde R_i, \tilde M, \tilde M_j \; (i=1,2,3,4, j=1,2)$ are given by 
\begin{equation}
\tilde L_i, \tilde R_i, \tilde M, \tilde M_j 
\equiv P\exp\{ig \int_{L_i, R_i, M, M_j} dx^{\mu}A_{\mu}(x)\} \in {\rm SU(3)}_c,
\end{equation}
{\it i.e.},  
$\tilde L_i, \tilde R_i, \tilde M, \tilde M_j\; (i=3,4, j=1,2)$ 
are line-like variables and 
$\tilde L_i, \tilde R_i\; (i=1,2)$ are staple-like variables, 
and $\tilde L_{12}, \tilde R_{12}$ are defined by  
\begin{equation}
\tilde L_{12}^{a'a}\equiv \frac12 \epsilon^{abc} \epsilon^{a'b'c'}
\tilde L_1^{bb'} \tilde L_2^{cc'},
\hspace{0.2cm}
\tilde R_{12}^{a'a}\equiv \frac12 \epsilon^{abc} \epsilon^{a'b'c'}
\tilde R_1^{bb'} \tilde R_2^{cc'}. 
\label{R12}
\end{equation}
Note that both the 4Q Wilson loop $W_{\rm 4Q}$ and the 5Q Wilson loop $W_{\rm 5Q}$ are gauge invariant.

\begin{figure}[ht]
\begin{tabular}{cc}
\hspace{-0.35cm}
\begin{minipage}{0.48\hsize}
\centerline{\epsfxsize=5.5cm\epsfbox{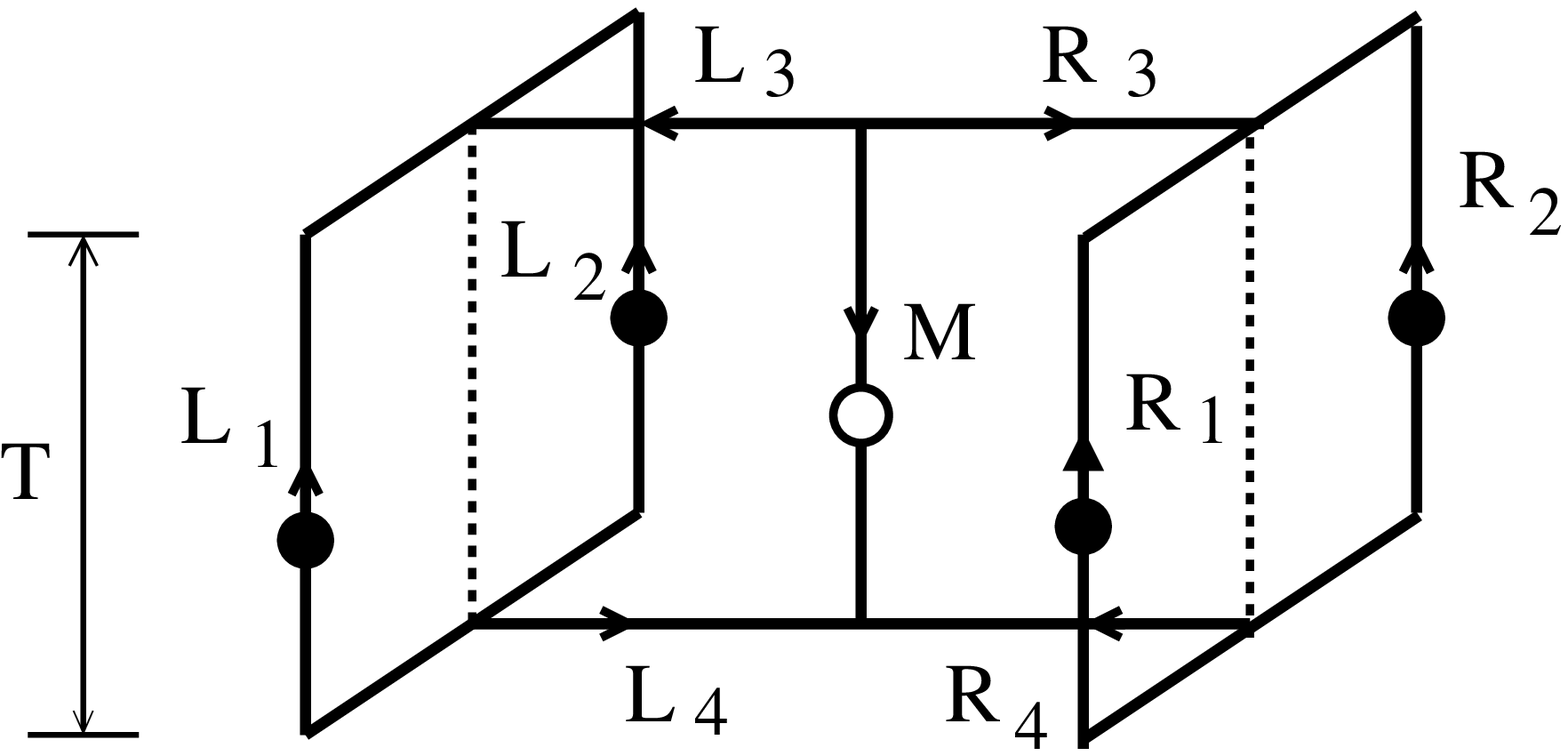}}   
\caption{The penta-quark (5Q) Wilson loop $W_{\rm 5Q}$. 
A gauge-invariant 5Q state is generated at $t=0$ and annihilated at $t=T$. 
Four quarks and an antiquark are spatially fixed for $0<t<T$.
}
\label{Fig1}
\end{minipage}
\hspace{0.2cm}
\begin{minipage}{0.48\hsize}
\centerline{\epsfxsize=4.8cm\epsfbox{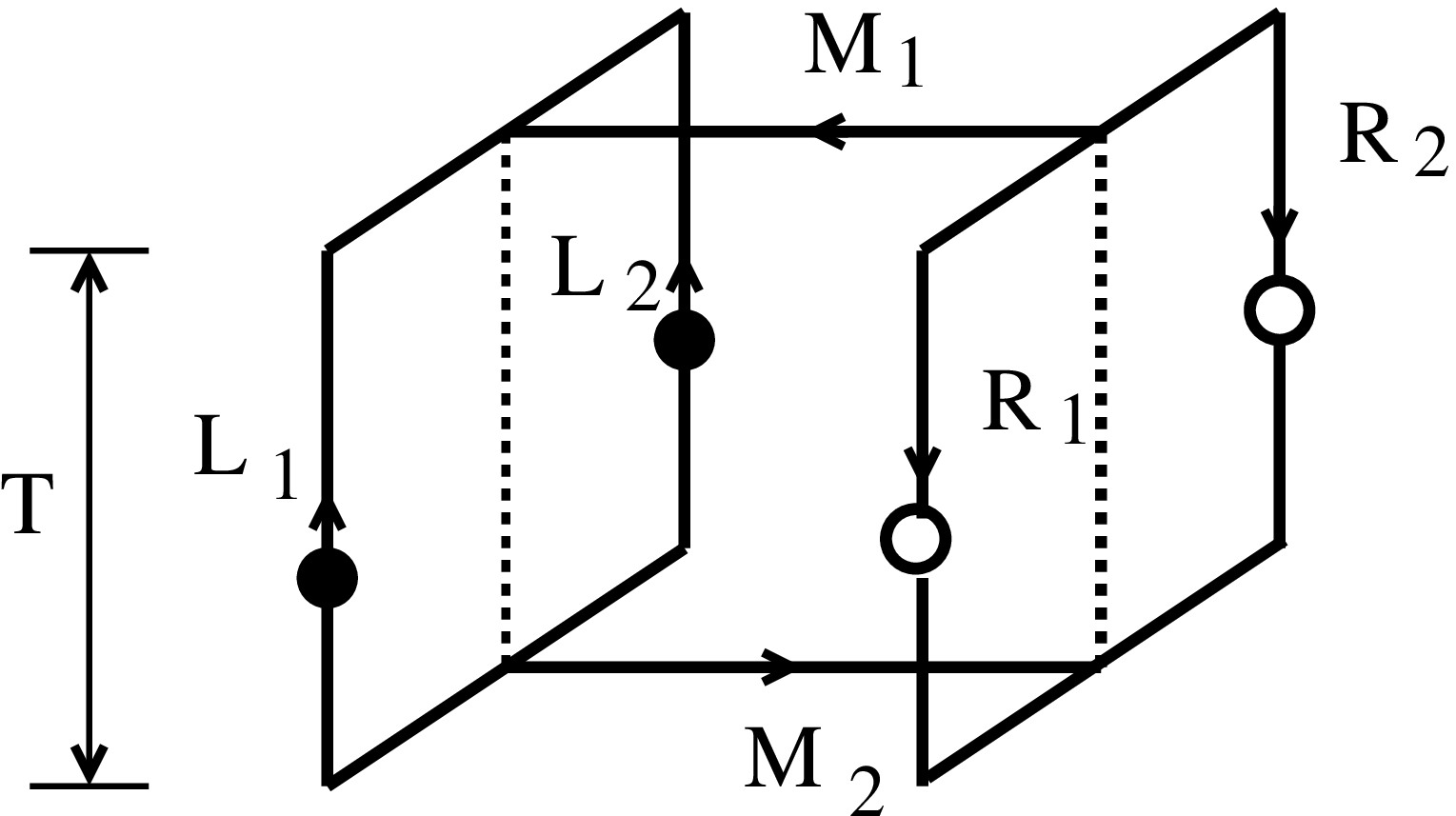}}   
\caption{The tetra-quark (4Q) Wilson loop $W_{\rm 4Q}$. 
A gauge invariant 4Q state is generated at $t=0$ and annihilated at $t=T$. 
Two quarks and two antiquarks are spatially fixed for $0<t<T$.
}
\label{Fig2}
\end{minipage}
\end{tabular}
\end{figure}

In general, the multi-quark operator in the multi-quark Wilson loop contains excited-state components. 
In order to extract the ground-state potential, 
we use the smearing method.\cite{TS} 
By this procedure, we obtain the quasi-ground-state operators for the static multi-quark systems, 
and thus perform the accurate calculations for the multi-quark potentials with them.

\section{Lattice QCD results and Concluding Remarks}

The lattice QCD simulation is performed at $\beta=6.0$, {\it i.e.}, $a \simeq 0.1~{\rm fm}$, 
on the $16^3 \times 32$ lattice at the quenched level.\cite{OST04,STOI04} 
In this paper, we investigate the planar and twisted configurations 
for the 5Q system as shown in Figs.3 and 4, 
and demonstrate the case of $d_1=d_2=d_3=d_4 \equiv d$ and $h_1=h_2 \equiv h/2$.
\begin{figure}[ht]
\vspace{-0.4cm}
\begin{tabular}{cc}
\hspace{-0.3cm}
\begin{minipage}{0.48\hsize}
\centerline{\epsfxsize=4cm\epsfbox{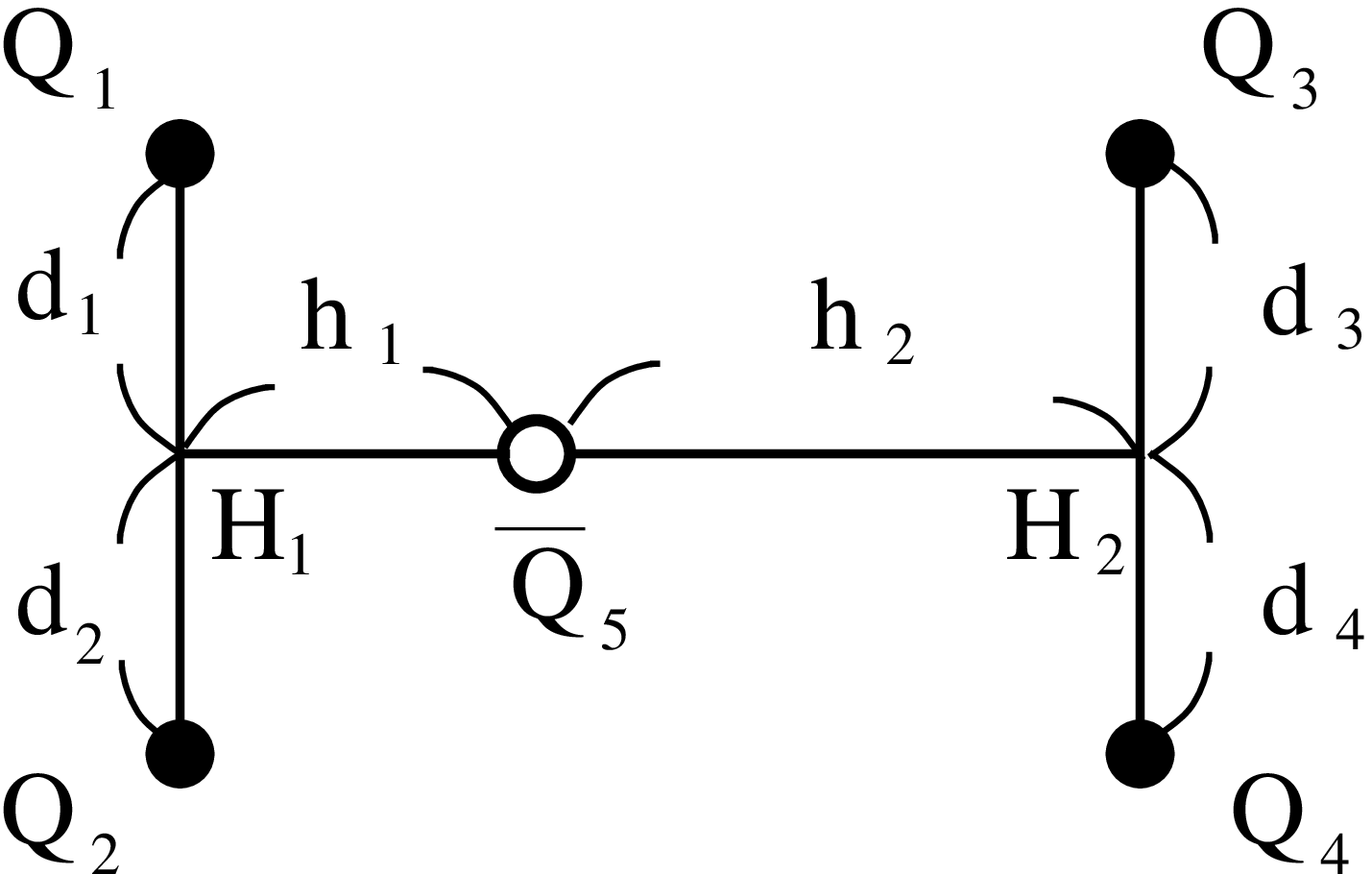}}
\caption{A planar configuration of the penta-quark system. 
\label{Fig3}}
\end{minipage}
\hspace{0.2cm}
\begin{minipage}{0.48\hsize}
\centerline{\epsfxsize=4.5cm\epsfbox{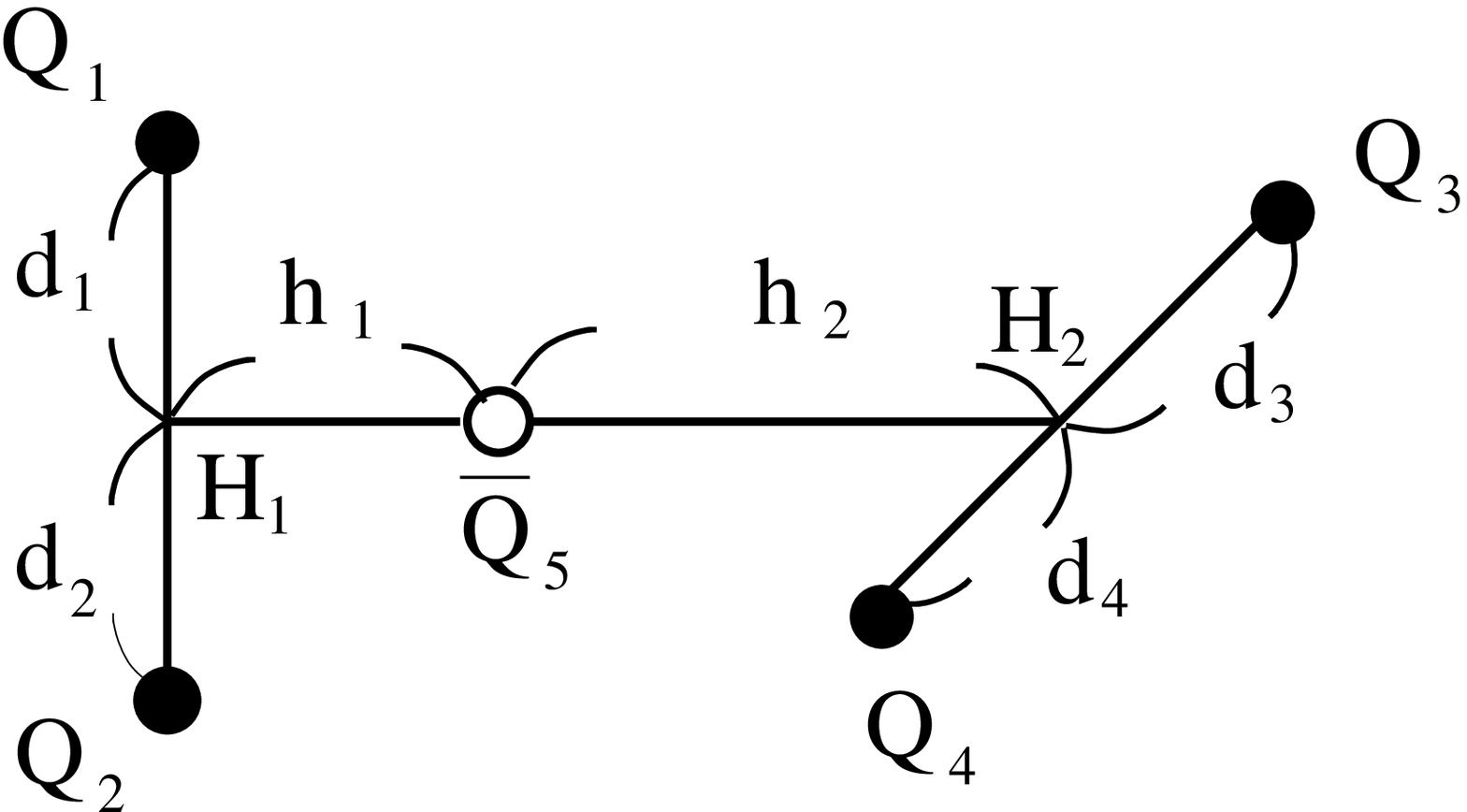}}
\caption{A twisted configuration of the penta-quark system with  
${\rm Q}_1{\rm Q}_2 \perp {\rm Q}_3{\rm Q}_4$.
\label{Fig4}}
\end{minipage}
\end{tabular}
\vspace{-0.36cm}
\end{figure}

In Fig.5, we show the lattice QCD results of the 5Q potential $V_{\rm 5Q}$. 
The lattice data denoted by the symbols are found to be well reproduced by 
the theoretical curves of the OGE plus multi-Y Ansatz\cite{OST04,STOI04}\footnote{
For the extreme case, {\it e.g.}, $d> \sqrt{3} h_1$, we here assume that the flux-tube is formed on the straight lines of   
${\rm Q}_1{\rm Q}_5$ and ${\rm Q}_2{\rm Q}_5$.} 
with $(A_{\rm 5Q},\sigma_{\rm 5Q})$ fixed to be $(A_{\rm 3Q},\sigma_{\rm 3Q}) \simeq (0.1366, 0.046a^{-2})$ 
in the 3Q potential.\cite{TS}\footnote{Due to this fixing, there is no adjustable parameter except for an irrelevant constant.}

\begin{figure}[h]
\vspace{-0.4cm}
\begin{center}
\includegraphics[height=1.51in]{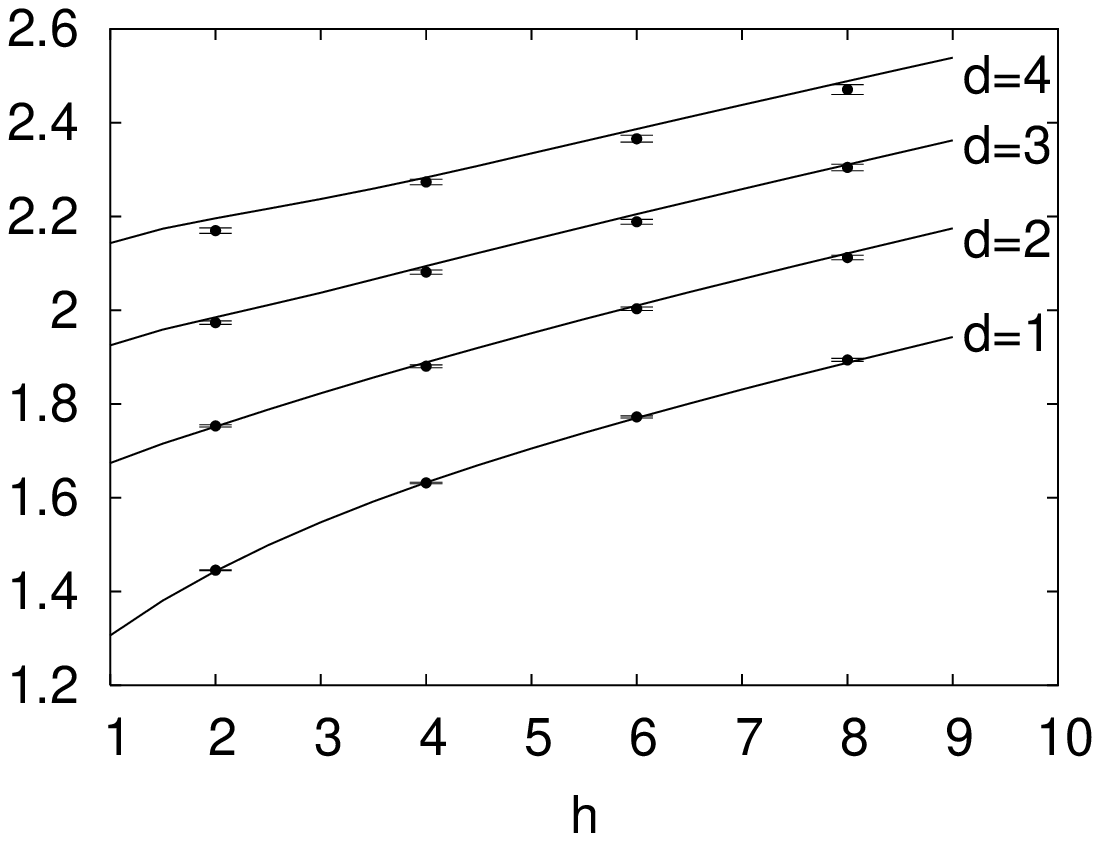}
\hspace{0.2cm} 
\includegraphics[height=1.51in]{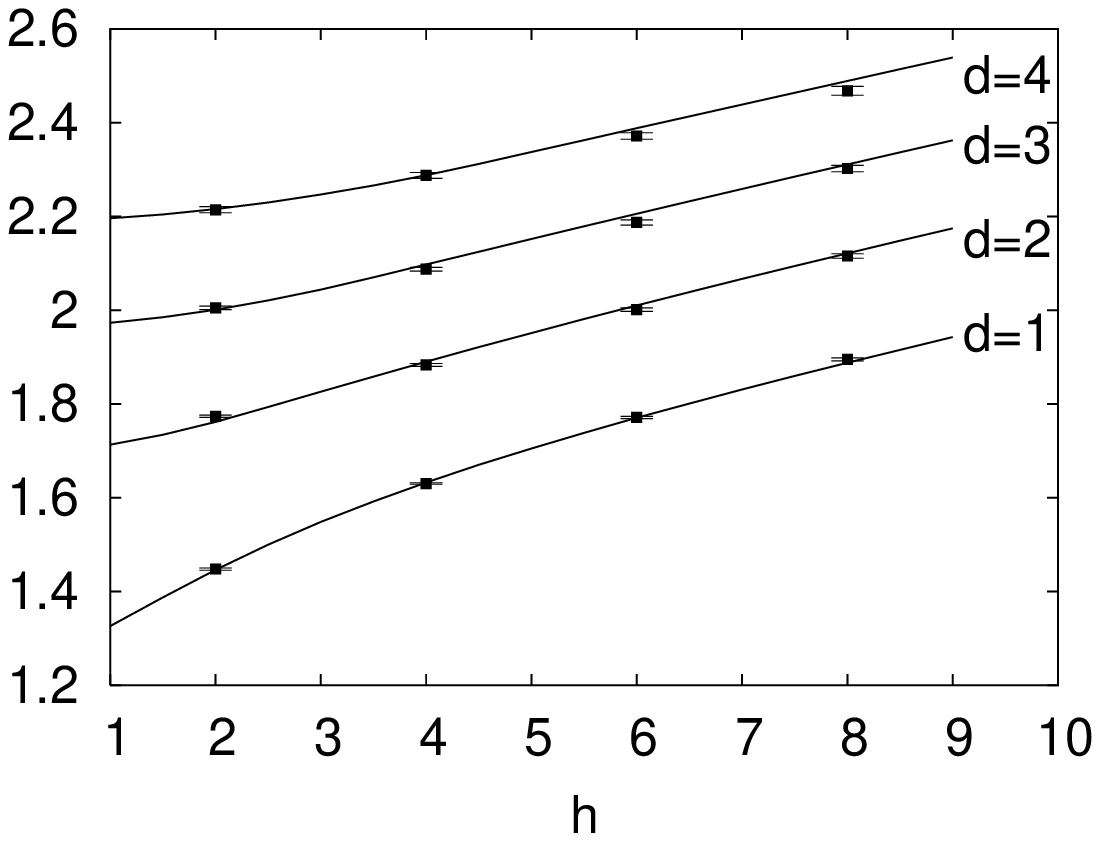} 
\vspace{-0.6cm}
\caption{Lattice QCD results of the 5Q potential $V_{\rm 5Q}$ in the lattice unit 
for the planar configurations (left) and the twisted configurations (right). 
The symbols denote the lattice data. The theoretical curves 
of the OGE plus multi-Y Ansatz are added. 
\label{Fig5}}
\vspace{-0.55cm}
\end{center}
\end{figure}

\begin{figure}[h]
\vspace{-0.5cm}
\begin{center}
\includegraphics[height=1.51in]{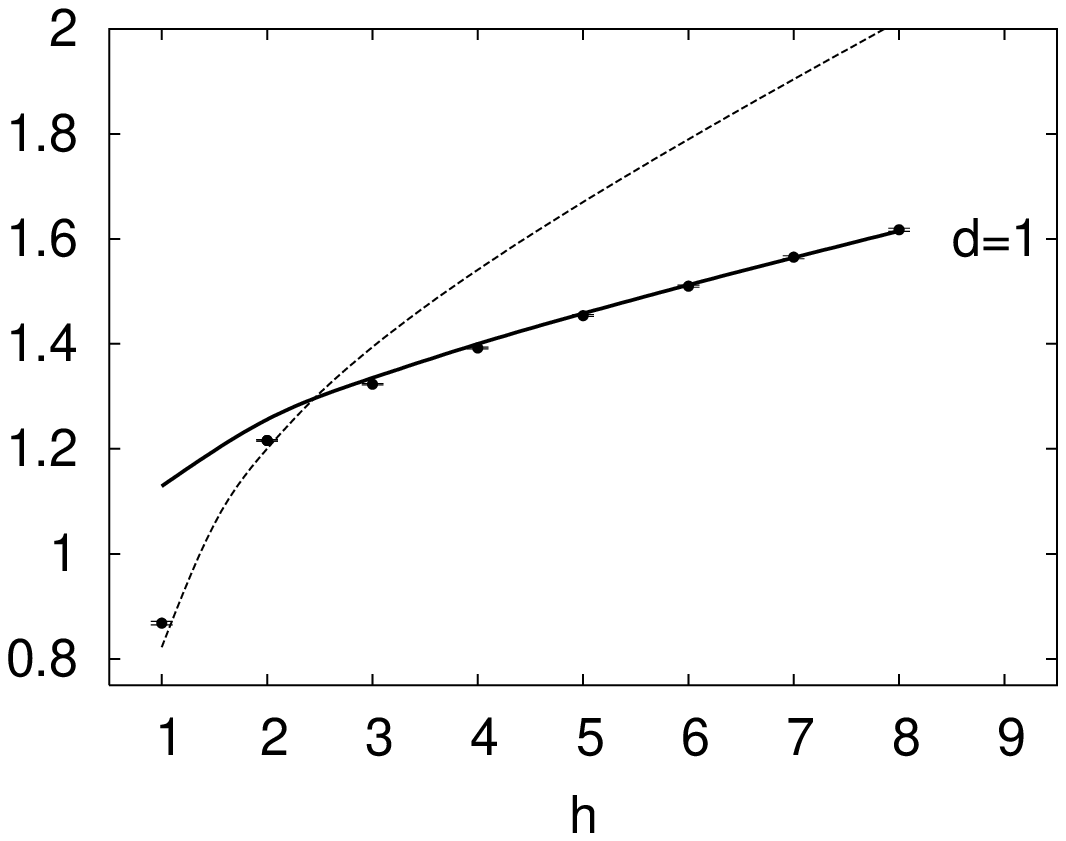}
\hspace{0.2cm} 
\includegraphics[height=1.51in]{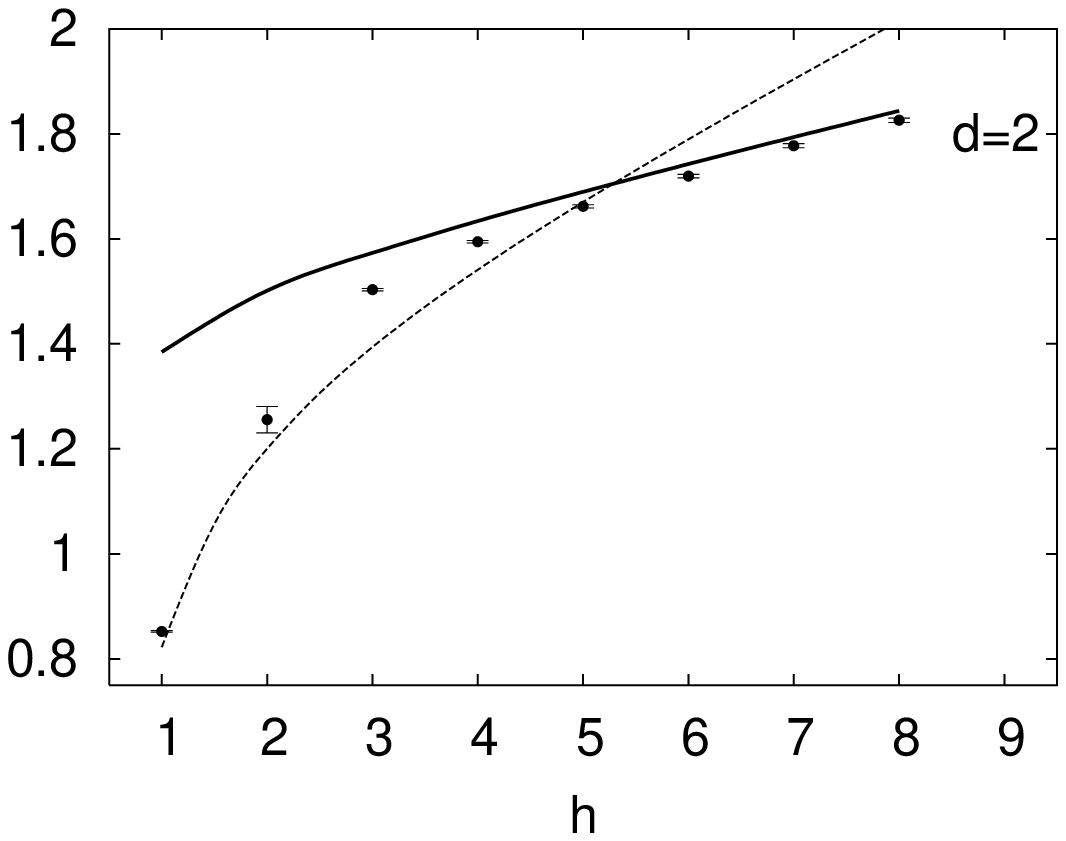} 
\vspace{-0.7cm}
\caption{The 4Q potential $V_{\rm 4Q}$ for $d=1$ (left) and $d=2$ (right) for the 
planar 4Q configuration similar to Fig.3. The horizontal axis $h$ corresponds to $h_1+h_2$.
The symbols denote the lattice QCD results.
The theoretical curves are added for the connected 4Q system (the solid curve) 
and for the ``two-meson" system (the dashed curve). 
\label{Fig6}}
\vspace{-0.4cm}
\end{center}
\end{figure}

Figure 6 shows the lattice results of the 4Q potential $V_{\rm 4Q}$.
For large $h$, $V_{\rm 4Q}$ coincides with the energy $V_{\rm c4Q}$ of the connected 4Q system. 
For small~$h$, $V_{\rm 4Q}$ coincides with the energy $2V_{\rm Q\bar Q}$ of the ``two-meson" system composed of two flux-tubes.
Thus, we get the relation of $V_{\rm 4Q} = {\rm min}  (V_{\rm c4Q}, 2V_{\rm Q\bar Q})$, 
and find the ``flip-flop" between the connected 4Q system and the ``two-meson" system around 
the level-crossing point where these two systems are degenerate as $V_{\rm c4Q}=2V_{\rm Q\bar Q}$. 
This result also indicates the flux-tube picture.

To summarize, we have performed the static 5Q and 4Q potentials in lattice QCD, and have found that  
the multi-quark potentials are well described with the OGE Coulomb plus multi-Y linear potential except for extreme cases. 
For the static 4Q potential, we have found the ``flip-flop" 
between the connected 4Q system and the ``two-meson" system. 
The present lattice QCD results for the multi-quark potentials 
provide a guiding principle in modeling the multi-quark system.


\vspace{-0.4cm}

\begin{thebibliography}{99}
\bibitem{Theta1540} LEPS Collaboration (T. Nakano {\it et al.}), 
{\it Phys. Rev. Lett.} {\bf 91}, 012002 (2003).
\bibitem{Z04} 
S.L.~Zhu, {\it Int. J. Mod. Phys.} {\bf A19}, 3439 (2004) and references therein.
\bibitem{NA49} 
NA49 Collaboration (C. Alt {\it et al.}), 
{\it Phys. Rev. Lett.} {\bf 92}, 042003 (2004). 
\bibitem{H1} H1 Collaboration (A.~Aktas {\it et al.}), 
{\it Phys. Lett.} {\bf B588}, 17 (2004).
\bibitem{DPP97} D.~Diakonov, V.~Petrov and M.~Polyakov, 
{\it Z. Phys.} {\bf A359}, 305 (1997).
\bibitem{JW03} R.L.~Jaffe and F.~Wilczek, 
{\it Phys. Rev. Lett.} {\bf 91}, 232003 (2003).
\bibitem{SR03} Fl. Stancu and D.O. Riska, {\it Phys. Lett.} {\bf B575}, 242 (2003). 
\bibitem{OST04} F. Okiharu, H. Suganuma and T.T. Takahashi, hep-lat/0407001 (2004).
\bibitem{STOI04}
H. Suganuma, T.T. Takahashi, F. Okiharu and H. Ichie, 
Proc. of {\it QCD Down Under}, March 2004, Adelaide, 
{\it Nucl. Phys.} {\bf B} (Proc. Suppl.) (2004) in press.
\bibitem{KL03} M.~Karliner and H.J.~Lipkin, hep-ph/0307243 (2003). 
\bibitem{Ken04} N. Mathur {\it et al.}, 
hep-ph/0406196 (2004), and references therein.
\bibitem{NSTV04} 
I.M.~Narodetskii {\it et al.}, {\it Phys. Lett.} {\bf B578}, 318 (2004). 
\bibitem{H03} A. Hosaka, {\it Phys. Lett.} {\bf B571}, 55 (2003).
\bibitem{BM04} P. Bicudo and G.M. Marques, {\it Phys. Rev.} {\bf D69}, 011503 (2004).
\bibitem{SZ04} X.-C. Song and S.-L. Zhu, hep-ph/0403093 (2004). 
\bibitem{KMN04} 
Y. Kanada-Enyo, O. Morimatsu and T. Nishikawa, hep-ph/0404144 (2004).
\bibitem{O04} For a review, M. Oka, 
{\it Prog. Theor. Phys.} {\bf 111}, 1 (2004) and its references. 
\bibitem{TS} 
T.T.~Takahashi {\it et al.}, 
{\it Phys. Rev. Lett.}~{\bf 86}, 18 (2001); {\it Phys. Rev.} {\bf D65}, 114509 (2002); 
{\it Phys. Rev. Lett.} {\bf 90}, 182001 (2003); {\it Phys. Rev.} {\bf D70}, 074506 (2004).
\end{thebibliography}
\end{document}